\documentclass[twocolumn,aps,prl] {revtex4} %% 2-spaltig
\usepackage{graphicx}
\begin{document}
\pagestyle{empty} 
\title{Rubber friction and tire dynamics}
\author{  
B.N.J. Persson} 
\affiliation{IFF, FZ-J\"ulich, D-52428 J\"ulich, Germany}
\affiliation{www.MultiscaleConsulting}

\begin{abstract}
We propose a simple rubber friction law, which can be used, e.g., in models of 
tire (and vehicle) dynamics. The friction law is tested by comparing numerical results to the 
full rubber friction theory (B.N.J. Persson, J. Phys.: Condensed Matter {\bf 18}, 7789 (2006)). 
Good agreement is found between the two theories. 

We describe a two-dimensional (2D) tire model which combines the rubber friction model with a simple mass-spring
description of the tire body. The tire model is very flexible and can be used to
calculate accurate $\mu$-slip (and the self-aligning torque) curves 
for braking and cornering or combined motion (e.g., braking during cornering). We 
present numerical results which illustrate the theory. Simulations of
Anti-Blocking System (ABS) braking are performed using two simple control algorithms.
\end{abstract}
\maketitle

%%%%%%%%%%%%%% main text %%%%%%%%%%%%%%%%
%\begin{multicols}{2}

{\bf 1. Introduction}

Rubber friction is a topic of huge practical importance, e.g., for tires, rubber seals,
conveyor belts and 
syringes\cite{Book,Grosch,tire,Pacejka,Persson1,JPCM,P3,theory1,theory2,Westermann,theory3,theory4,theory5,theory6,wear,Mofidi,Creton}. 
% (between the rubber plunger and the syringe substrate).
In most theoretical studies rubber friction is described using very simple 
phenomenological models, e.g., the Coulombs friction law with a friction coefficient which
may depend on the local sliding velocity.
However, as we have shown earlier\cite{JPCM}, rubber friction depends on
the {\it history} of the sliding motion  (memory effects), which we have found to be crucial
for an accurate description of rubber friction. For rubber sliding on a hard rough substrate,
the history dependence of the friction is mainly due to frictional heating in the rubber-substrate
contact regions. Many experimental observations, such as an apparent dependence of the rubber friction
on the normal stress, can be attributed to the influence of frictional heating on the rubber friction. 

A huge number of papers have been published related to
tire dynamics, in particular in the context of Anti-Blocking System (ABS) braking models. The ``heart'' in
tire dynamics is the road-rubber tire friction. Thus, unless this friction is accurately described,
no tire model, independent of how detailed the description of the tire body may be, will provide
an accurate picture of tire dynamics. However, most treatments 
account for the road-tire friction in a very approximate way. Thus, many ``advanced''
finite element studies for tire dynamics account for the friction only via a static and a kinetic rubber
friction coefficients. In other studies the dynamics of the the whole tire is described using 
interpolation formulas, e.g., the ``Magic Formula''\cite{Pacejka}, 
but this approach require a very large set of 
measured tire properties (which are expensive and time-consuming to obtain), and cannot 
describe the influence of history (or memory) effects on tire dynamics.   

In this paper we first propose a very simple rubber friction law (with memory effects) which 
gives nearly identical results to the full model developed in Ref. \cite{JPCM}. 
We also develop a 2D-tire model which combines the rubber-road friction
theory (which accounts for the flash temperature) with a simple two-dimensional
(2D) description of the tire body. We believe that the most important aspect of the tire body is 
its distributed mass and elasticity, and this is fully accounted for in our model. One advantage of the
2D-model over a full 3D-model is that one can easily impose any foot-print pressure distribution
one like (e.g., measured pressure distributions), while in a 3D-model the pressure distribution
is fixed by the model itself. This allow a detailed study on how sensitive the tire dynamics depends
on the nature of the foot print pressure distribution. 
The tire model is illustrated by calculating $\mu$-slip curves, and with simulations of
ABS braking using two different control algorithms.

\begin{figure}
\includegraphics[width=0.45\textwidth,angle=0]{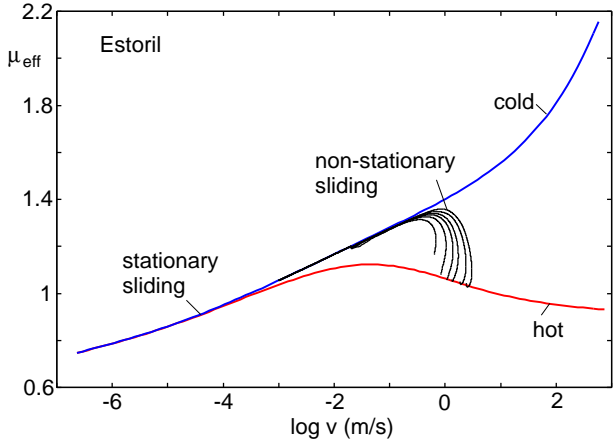}
\caption{\label{muslip}
Red and blue lines: the kinetic friction coefficient (stationary sliding) as a function of the logarithm 
(with 10 as basis) of the sliding velocity. The blue line denoted ``cold'' is without
the flash temperature while the red line denoted ``hot'' is with the flash temperature.
Black curves: the effective friction experienced by a tread block  as it goes through the foot print.
For the car velocity $27 \ {\rm m/s}$ and for several 
slip values $0.005$, $0.0075$, $0.01$, $0.03$, $0.05$, $0.07$ and $0.09$. Note that the friction
experienced by the tread block first follows the ``cold'' rubber branch of the steady state kinetic friction coefficient and then, when the block has slip a distance of order the diameter of the macroasperity contact region,
it follows the ``hot'' rubber branch. 
}
\end{figure}

\vskip 0.3cm
{\bf 2. Rubber friction}

Rubber friction depends on the history of the sliding motion. This is mainly due to
the flash temperature: the temperature in the rubber-road asperity contact regions at time
$t$ depends on the sliding history for all earlier times $t' < t$. This memory effect is crucial for an
accurate description of rubber friction. We illustrate this effect in Fig. \ref{muslip}
for a rubber tread block sliding on an asphalt road surface. 
We show the (calculated) kinetic friction coefficient for stationary sliding without (blue curve)
and including the flash temperature (red curve) as a function of the velocity $v$
of the bottom surface of the rubber block. The black curves shows the effective friction during
non-stationary sliding experienced by a rubber tread block during braking at various slips
(slip values from 0.005 to 0.09). Note that because some finite sliding distance is necessary in order
to fully develop the flash temperature, the friction acting on the tread block initially 
follows the blue curve corresponding to
``cold-rubber'' (i.e., negligible flash temperature). 
Thus, it is not possible to accurately describe rubber friction
with just a static and a kinetic friction coefficients (as is often done even in advanced tire 
dynamics computer
simulation codes) or even with a function $\mu (v)$ which depends on the instantaneous sliding velocity
$v(t)$. Instead, the friction depends on $v(t')$ for all times $t' \le t$.

\begin{figure}
\includegraphics[width=0.35\textwidth]{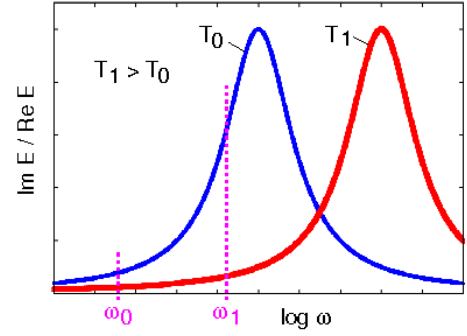} 
\caption{\label{tandd} 
When the temperature increases the ${\rm tan}\delta = {\rm Im}E/{\rm Re}E$ spectra shift
to higher frequencies, which result in a decrease in the rubber friction. We assume the 
road asperities gives rise to pulsating frequencies in the range $\omega_0$ and $\omega_1$.
}
\end{figure}

As a background to what follows,
we first review the rubber friction theory (see \cite{JPCM,Persson1} for details).
It is assumed that all energy dissipation arises from the viscoelastic deformations of the
rubber surface by the road asperities. An asperity contact region with the diameter $d$ gives rise to
time dependent (pulsating) deformations of the rubber which is characterized by the frequency $\omega = v/d$,
where $v$ is the sliding velocity. 
The viscoelastic deformation (and most of the energy dissipation) extend into the rubber by
the typical distance $d$. So most of the energy dissipation occur in a volume element of order $d^3$.
In order to have a large asperity-induced contribution to the friction the frequency $\omega$ should be
close to the maximum of the ${\rm tan}\delta = {\rm Im} E(\omega) /{\rm Re} E(\omega)$ curve. Here
$E(\omega )$ is the viscoelastic modulus of the rubber. In reality there will be a wide distribution 
of asperity contact sizes, so there will be a wide range of perturbing frequencies, say from $\omega_0$
to $\omega_1$, see Fig. \ref{tandd}. A large friction requires that ${\rm tan}\delta$ 
is as large as possible for all these perturbing frequencies.

The temperature dependence of the viscoelastic modulus 
of rubber-like materials is usually very strong,
and an increase in the temperature by $10 \ ^\circ {\rm C}$ may shift the ${\rm tan}\delta$
curve to higher frequencies by one frequency decade. This will usually reduce the rubber friction, see Fig. \ref{tandd}.

\begin{figure}
\includegraphics[width=0.43\textwidth]{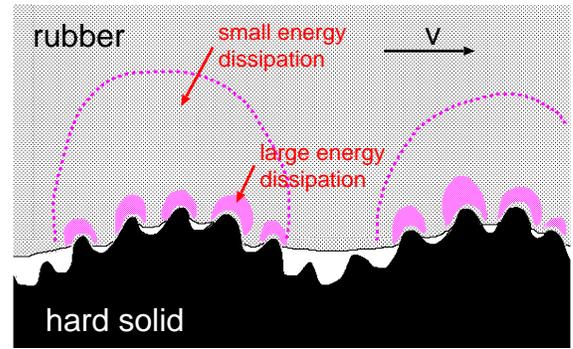} 
\caption{\label{twolengthscales} 
The dissipated energy per unit volume is highest in the smallest asperity
contact regions.
}
\end{figure}

Real surfaces have a wide distribution of asperity sizes. 
The best picture of a rough surface is to think about it as
big asperities on top of which occur smaller asperities on top of which
occur even smaller asperities ... . This is illustrated in Fig. \ref{twolengthscales} for a case where
roughness occur on two length scales. To get the total energy dissipation during sliding on
a real surface one need to sum up the contribution from asperity induced deformation of the
rubber on all (relevant) length scales. It is important to note that all length scales are a priory
equally important\cite{Persson1}. 

Temperature has a crucial influence on rubber friction. The viscoelastic
energy dissipation, which is the origin of the rubber friction in my model, result in local heating
of the rubber in exactly the region where the energy dissipation occur. 
This result in a temperature increase, which becomes larger 
as we observe smaller and smaller asperity contact regions. This local (in time and space) 
temperature increase, resulting from the local viscoelastic energy dissipation, is referred to as the
{\it flash temperature}. The flash temperature 
has an extremely important influence on the rubber friction. This is illustrate
in Fig. \ref{muslip}, where we show the (calculated)
steady state kinetic friction coefficient when a block of tread rubber
is sliding on an asphalt road surface. 
The upper curve is the result without accounting for the flash temperature, i.e., the 
temperature is assumed to be equal to the background temperature $T_0$ everywhere. 
The lower curve is the result including the flash temperature. Note that for 
sliding velocities $v > 0.001 \ {\rm m/s}$ the flash temperature result in a lowering of the sliding
friction. For velocities $v < 0.001 \ {\rm m/s}$ the produced heat has enough time to diffuse away from the 
asperity contact regions, and the flash temperature effect is negligible. 

\begin{figure}
\includegraphics[width=0.35\textwidth]{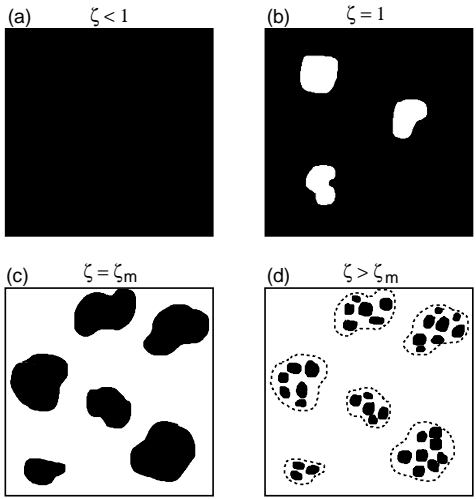} 
\caption{\label{PictureMag} 
The contact region between a tire an a road surface. At low magnification $\zeta < 1$
it appears that the tire is in complete contact with the road but as the
magnification increases, the contact area continuously decreases as indicated in the figure.
}
\end{figure}

In the rubber friction theory the concept of the {\it macroasperity} 
contact region is of crucial importance. Let us
study the footprint contact region between a tire and a road surface at different magnification $\zeta$.
At low magnification the road surface appears smooth and the contact between the tire and the road 
appears to be complete within the footprint area as in Fig. \ref{PictureMag}(a). However, when we increase
the magnification $\zeta$ we start to observe non-contact regions as in Fig. \ref{PictureMag}(b). 
At high enough magnification we observe isolated contact regions as in Fig. \ref{PictureMag}(c),
which, when the magnification increases even further, 
break up into even smaller contact regions as in Fig. \ref{PictureMag}(d). 
We denote the contact regions observed in Fig. \ref{PictureMag}(c) as the {\it macroasperity} 
contact regions (with the average diameter $D$) (the exact definition is given in Ref. \cite{JPCM,PSSR}).
If the nominal pressure in the tire-road contact region is small the macroasperity contact regions 
will be well separated, but the separation between the {\it microasperity} contact regions within
the macroasperity contact regions is in general very small. When calculating the flash temperature
effect we have therefore smeared out the heat produced by the microasperity contact regions uniformly
within the macroasperity contact regions. 
Typically for road surfaces $D\approx 0.1-1
\ {\rm cm}$, and the fraction of the tread block surface occupied
by the macroasperity contact regions are typically between $10\%$ and $30\%$.

In the friction theory developed in Ref. \cite{JPCM} only the surface roughness with wave vectors
$q < q_1$ are assumed to contribute to the friction. For clean road surfaces we determine the cut-off
wavevector $q_1$ by a yield condition: We assume that the local stress and temperature in the asperity contact regions
on the length scale $1/q_1$ are so high that the rubber bonds break resulting in a thin modified (dead) layer of rubber
at the surface region of thickness $\approx 1/q_1$. From this follows the following observations:

(a) The rubber friction on clean road surfaces after run-in is rather insensitive to the road 
surface. This has been observed in several series of experimental studies (not shown), and is in accordance
with the present theory. This can be understood as follows. The cut-off $q_1$ on surfaces with
smoother, less sharp roughness or surfaces where the roughness occur at shorter length scales
will be larger (i.e. the cut-off wavelength $\lambda_1 = 2\pi /q_1$ smaller) than for road surfaces
with larger roughness in such a way as the stress and temperature increase in the asperity contact regions
which can be observed at the resolution $\lambda_1$ (or magnification $\zeta = q_1/q_0$) are roughly the
same on all surfaces. This implies that a larger range of roughness will contribute to the rubber
friction on ``smoother'' surfaces as compared to more rough surfaces. 
As a consequence, the friction (after run-in) may vary very little between
different (clean) road surfaces. 

(b) On contaminated road surfaces, the cut-off $q_1$ may be determined by the nature of the contamination.
In this case, if the cut-off is fixed (e.g., determined by, say, the size of contamination particles) 
one may expect much larger variation in the friction coefficient between different road surfaces, and also
a larger variation between tired with different types of tread rubber.

\vskip 0.3cm
{\bf 3. Phenomenological rubber friction law}

In tire applications, for slip of order $5-10\%$ and typical footprint length of order $10 \ {\rm cm}$,
the slip distance of a tread rubber block in the footprint will be of order $1 \ {\rm cm}$,
which typically is of order the diameters $D$
of the macro asperity contact regions. As discussed above, as long as the slip distance 
$s(t)$ is small compared to $D$ one follow the cold rubber branch of the steady state
relation $\mu (v)$ so that $\mu(t) \approx \mu_{\rm cold}(v(t))$ for the slip distance 
$s(t) << D$. When the tread block moves towards the end of the footprint the slip distance $s(t)$
may be of order of (or larger than) $D$, and the friction will follow the hot branch of the 
$\mu (v)$ relation i.e., $\mu(t) \approx \mu_{\rm hot}(v(t))$ for $s(t) > D$. 
We have found that the following (history dependent) friction law gives nearly the same result as the full
theory presented in Ref. \cite{JPCM}:
$$\mu_{\rm eff} (t) = \mu_{\rm cold} (v(t),T_0) e^{-s(t)/s_0}$$
$$+\mu_{\rm hot} (v(t),T_0) 
\left [1-e^{-s(t)/s_0}\right ]\eqno(1)$$
where $v(t)$ is the instantaneous sliding velocity and $s(t)$ 
the sliding distance, and $s_0 \approx 0.2 D$. We will refer to (1) as the {\it cold-hot friction law}.
The length $D$ depends on the rubber compound and the road surface but is typically in the range 
$D\approx 0.1-1 \ {\rm cm}$. Using the full friction theory one can easily calculate the functions
$\mu_{\rm cold} (v(t),T_0)$ and $\mu_{\rm hot} (v(t),T_0)$ and the length $D$. 

%\begin{figure}
%\includegraphics[width=0.45\textwidth,angle=0]{normal.stress.distance.ps}
%\caption{\label{normal.stress.distance}
%The normal pressure distribution in the foot print.
%}
%\end{figure}

\begin{figure}
\includegraphics[width=0.45\textwidth,angle=0]{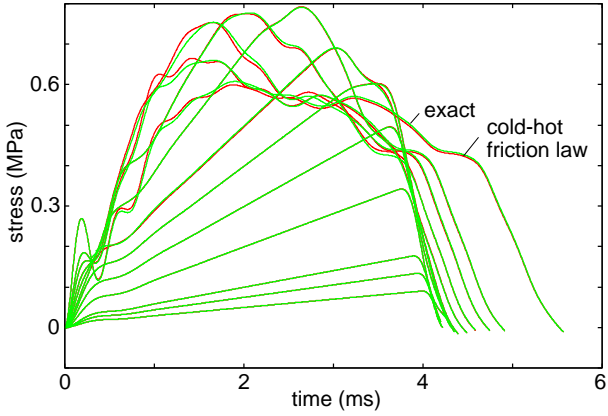}
\caption{\label{stress.time}
The frictional shear stress acting on a tread block as a function of time for many slip values: 
 $0.005$, $0.0075$, $0.01$, $0.03$, $0.05$, $0.07$, $0.09$, $0.12$, $0.15$ and $0.25$.
For the car velocity $27 \ {\rm m/s}$ and tire background temperature $T_0=60 \ ^\circ {\rm C}$.
For the 1D tire model using the
full friction model (green curves) and the cold-hot friction law (1) (red curves).
For a passenger car tread compound.
}
\end{figure}

\begin{figure}
\includegraphics[width=0.45\textwidth,angle=0]{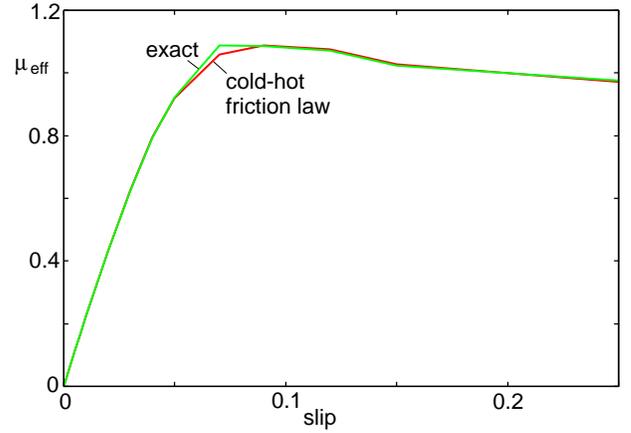}
\caption{\label{mu.slip}
The $\mu$-slip curve for the 1D tire model using the
full friction model (green curve) and the cold-hot friction law (1) (red curve).
For a passenger car tread compound.
}
\end{figure}

To demonstrate the accuracy of the cold-hot rubber friction law (1), let us 
study the dynamics of one tread block as it pass through the tire-road footprint.
%, using a 1D tire model.
%The tire-road footprint pressure distribution is shown in Fig. \ref{normal.stress.distance}. 
In Fig. \ref{stress.time} we show the 
the frictional shear stress acting on a tread block as a function of time for many slip values:
$0.005$, $0.0075$, $0.01$, $0.03$, $0.05$, $0.07$, $0.09$, $0.12$, $0.15$ and $0.25$.
Note that the 
cold-hot friction law (1) (red curves) gives nearly the same result as for the
full friction model (green curves).
In Fig. \ref{mu.slip} we show the $\mu$-slip curve. Again the cold-hot friction law (1) (red curve)
gives nearly the same result as the full friction model (green curve).

\vskip 0.3cm
{\bf 4. Tire dynamics}

All the calculations presented in this section have been obtained using the 2D tire model described below, 
with the tire body optimized using experimental data for a passenger car tire. The viscoelastic springs associated with this
tire body are kept fixed in all the calculations. Thus the model calculations does not take into
account the changes in the tire body viscoelastic properties
due to variations in the  tire (background) temperature, 
or variations in the tire inflation pressure (which affect the tension in the tire walls). 
In principle both effects can be relatively simply accounted for in the model, but have not been included so far.
All the calculations presented below are obtained using the measured viscoelastic modulus of a
passenger car tire tread compound, and for the Estoril racer track.

\begin{figure}
\includegraphics[width=0.45\textwidth]{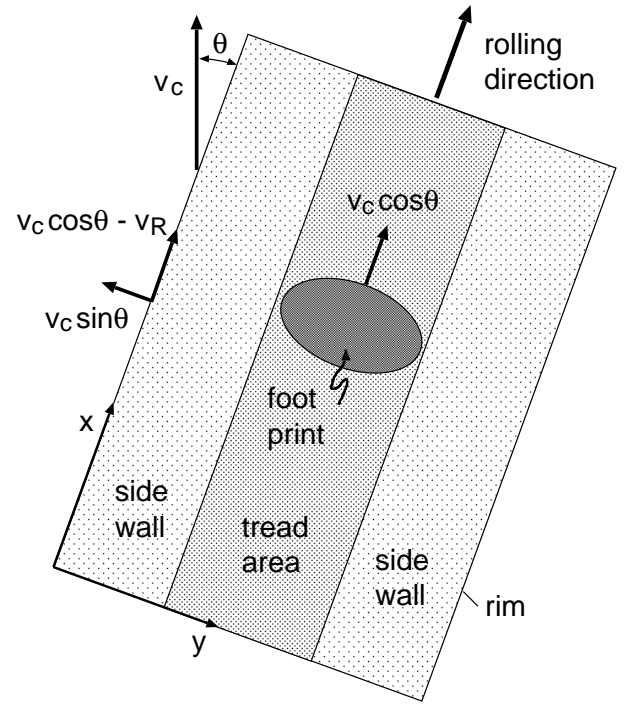}
\caption{\label{Tire} 
2D-model of a tire. The car velocity ${\bf v}_c$ points in
another direction than the rolling direction, giving a non-zero
cornering angle $\theta$.} 
\end{figure}

We use a 2D-description of the tire body as indicated in Fig. \ref{Tire}. 
We introduce a coordinate system with the $y$-axis in the transverse direction
and the $x$-axis along the longitudinal (rolling) direction. We consider the
tire-road system in a reference frame where the road is stationary. The
car velocity $v_{\rm c}$, the rolling velocity $v_{\rm R}$ and the cornering angle $\theta$
determines the transverse $v_y$ and longitudinal $v_x$ slip velocities: 
$$v_y=v_{\rm c} {\rm sin}\theta$$
$$v_x=v_{\rm c} {\rm cos}\theta - v_{\rm R}$$
The longitudinal slip is defined by
$$s= {v_x \over v_{\rm c} {\rm cos}\theta} = {v_{\rm c} {\rm cos}\theta - v_{\rm R} 
\over v_{\rm c} {\rm cos}\theta}$$
When the cornering angle $\theta = 0$ this equation reduces to
$$s= {v_{\rm c} - v_{\rm R} \over v_{\rm c}}$$
Note that $v_x$ and $v_y$ are also the velocities of the tire rim, and that the
footprint moves (in the rolling direction) 
with the velocity $v_{\rm c} {\rm cos}\theta$ relative to the road, and with the
rolling velocity $v_{\rm c} {\rm cos}\theta - v_x = v_{\rm R}$ relative to the rim.

We describe the tire body as a set of mass points connected with viscoelastic
springs (elasticity $k$ and viscous damping $\gamma$).
The springs have both elongation and bending elasticity, denoted by $k$ and $k'$, respectively,
and the corresponding viscous damping coefficients $\gamma$ and $\gamma'$.
We assume $N_x$ and $N_y$ tire body blocks along the $x$- and $y$-directions and let
${\bf x}_{ij}=(x_{ij},y_{ij})$ 
denote the displacement vector of tire body block $(i,j)$ ($i=1,...,N_x$, $j=1,...,N_y$).
Since the tire is a torus shaped object we must assume periodic boundary conditions in
the $x$-direction so that $x_{N_x+1,j}=x_{1,j}$ and $y_{N_x+1,j}=y_{1,j}$. 

For stationary tire motion we have the following boundary conditions.
For $i=0,...,N_x+1$:
$$y_{i0}=v_yt, \ \ \ \ \ x_{i0}=v_xt$$
$$y_{i,N_y+1}=v_yt, \ \ \ \ \ x_{i,N_y+1}=v_xt$$
For $j=1,...,N_y$:
$$y_{N_x+1,j}=y_{1j}, \ \ \ \ \   
x_{N_x+1,j}=x_{1j}$$
$$y_{0j}=y_{N_x,j}, \ \ \ \ \   
x_{0j}=x_{N_x,j}$$

If the mass
of tire body element $(i,j)$ is denoted by $m_{j}$, we get  
for $i=1,...,N_x$, $j=1,...,N_y$:
$$m_j \ddot y_{ij} = F_{yij}+k_{yj} (y_{i,j-1,}-y_{ij})
+k_{yj+1}(y_{i,j+1}-y_{ij})$$  
$$+\gamma_{yj} (\dot y_{i,j-1}-\dot y_{ij})+\gamma_{yj+1}(\dot y_{i,j+1}-\dot y_{ij})$$  
$$+k'_{xj} (y_{i+1,j}+y_{i-1,j}-2y_{ij})$$
$$
+\gamma'_{xj} (\dot y_{i+1,j}+\dot y_{i-1,j}-2\dot y_{ij})$$

$$m_j \ddot x_{ij} = F_{xij}+k_{xj} (x_{i-1,j}-x_{ij})+k_{xj}(x_{i+1,j}-x_{ij})$$  
$$+\gamma_{xj} (\dot x_{i-1,j}-\dot x_{ij})+\gamma_{xj}(\dot x_{i+1,j}-\dot x_{ij})$$  
$$+k'_{yj} (x_{i,j-1}-x_{ij})
+k'_{yj+1} (x_{i,j+1}-x_{ij})$$
$$+ \gamma'_{yj} (\dot x_{i,j-1}-\dot x_{ij})
+\gamma'_{yj+1} (\dot x_{i,j+1}-\dot x_{ij})$$

In the equations above, $F_{xij}$ and $F_{yij}$ are the force components 
(in the $x$- and $y$-directions,
respectively) 
acting on the tire body block
$(i,j)$ from the tread block $(i,j)$. Thus, ${\bf F}_{ij} = (F_{xij},F_{yij})$ 
is nonzero only when $(i,j)$ is in the
tire tread area.
The tire body viscoelastic spring parameters $(k,\gamma)$ and $(k',\gamma')$ in the tread area
and in the side wall area (8 parameters) have been optimized in order to reproduce a number of measured
tire properties (e.g., the longitudinal and transverse 
tire stiffness values for three tire loads, and the frequency and damping of the
lowest longitudinal and transverse tire vibrational modes).
The optimization has been performed using the amoeba method of multidimensional minimization\cite{amoeba}.

\begin{figure}
\includegraphics[width=0.4\textwidth,angle=0.0]{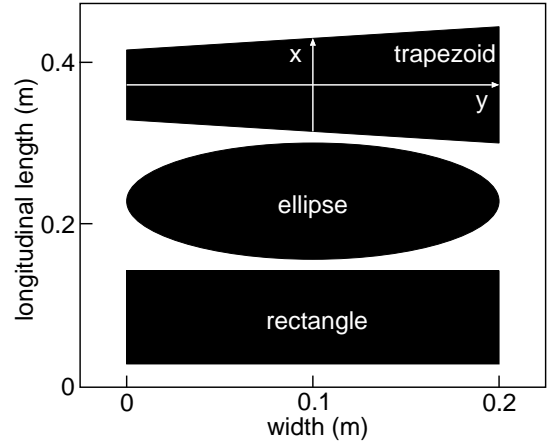}
\caption{\label{footprint}
Three tire-road footprints used in the computer simulations. The footprints
are $0.2 \ {\rm m}$ wide and the normal pressure in the footprints is
constant at $p = 0.1 \ {\rm MPa}$ or $0.3 \ {\rm MPa}$. 
}
\end{figure}

\vskip 0.2cm
{\bf 4.1 Dependence of the $\mu$-slip curve on the shape of the tire-road footprint}

Here we will study how the tire dynamics depends on the shape of the tire-road footprint.
We consider the rectangular, elliptic and trapezoid footprints shown
Fig. \ref{footprint}. 
We assume first that the pressure in the footprint is uniform and equal to $p= 0.1 \ {\rm MPa}$,
and that the tire load is equal to $2000 \ {\rm N}$ in all cases. Thus all the foot prints have the same area.

\begin{figure}
\includegraphics[width=0.45\textwidth,angle=0]{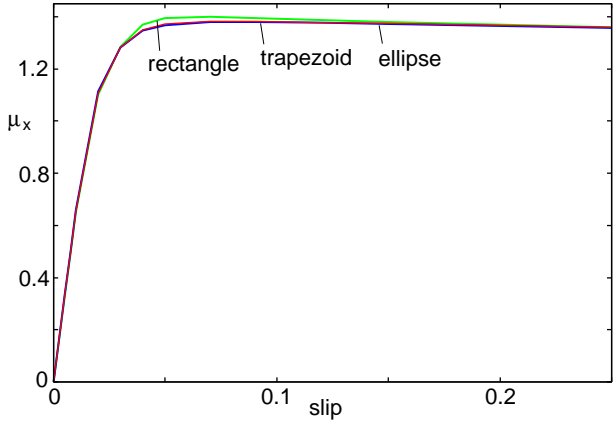}
\caption{\label{logv.Mueff.RectangleElipsTapetz}
The $\mu$-slip curves for the three footprints shown 
in Fig. \ref{footprint}. For the rubber background temperature $T_0=80 \ ^\circ {\rm C}$
and the car velocity $27 \ {\rm m/s}$. 
For $F_{\rm N} = 2000 \ {\rm N}$ and $p= 0.1 \ {\rm MPa}$.  
}
\end{figure}

\begin{figure}
\includegraphics[width=0.45\textwidth,angle=0]{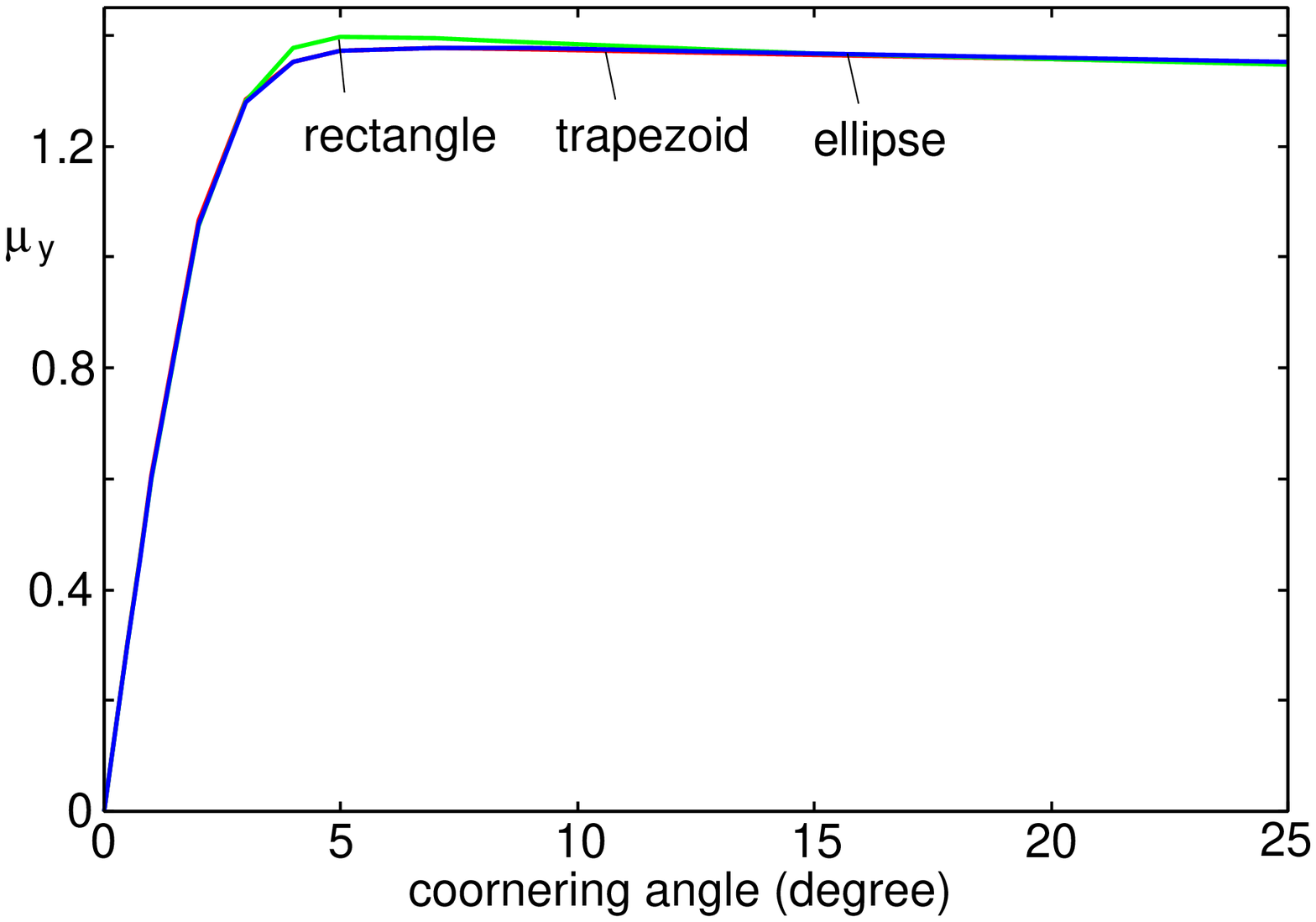}
\caption{\label{mu.slipangle.elips.rect.trapetz}
The $\mu$-slip angle curves for the three footprints shown
in Fig. \ref{footprint}. For the rubber background temperature $T_0=80 \ ^\circ {\rm C}$
and the car velocity $27 \ {\rm m/s}$. 
For $F_{\rm N} = 2000 \ {\rm N}$ and $p = 0.1 \ {\rm MPa}$.  
}
\end{figure}

In Fig. \ref{logv.Mueff.RectangleElipsTapetz} we show the 
$\mu$-slip curves for the three footprints shown
in Fig. \ref{footprint}, and in Fig. \ref{mu.slipangle.elips.rect.trapetz} we show
the corresponding $\mu$-slip angle curves. 
It is remarkable how insensitive the results are to the shape of the footprint.
Thus we may state that {\it $\mu$-slip curves, and
hence tire dynamics, depends very weakly on the shape of the tire-road 
footprint}, assuming everything else the same.

\begin{figure}
\includegraphics[width=0.45\textwidth,angle=0]{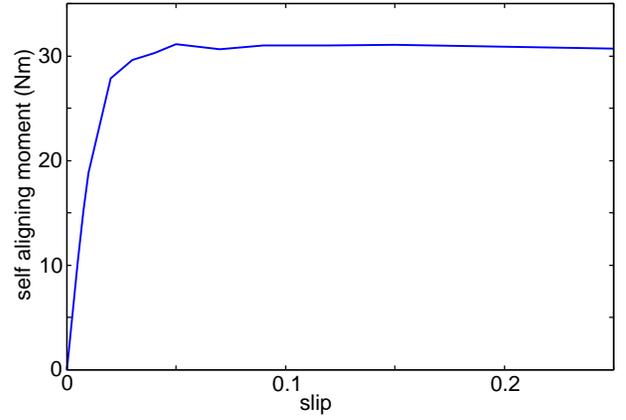}
\caption{\label{selfAligningMoment.slip.Trapetz}
The self-aligning moment for the trapezoid footprint, as a function of the longitudinal slip. 
For the rubber background temperature $T_0=80 \ ^\circ {\rm C}$
and the car velocity $27 \ {\rm m/s}$. 
For $F_{\rm N} = 2000 \ {\rm N}$ and $p = 0.1 \ {\rm MPa}$.  
}
\end{figure}

\begin{figure}
\includegraphics[width=0.45\textwidth,angle=0]{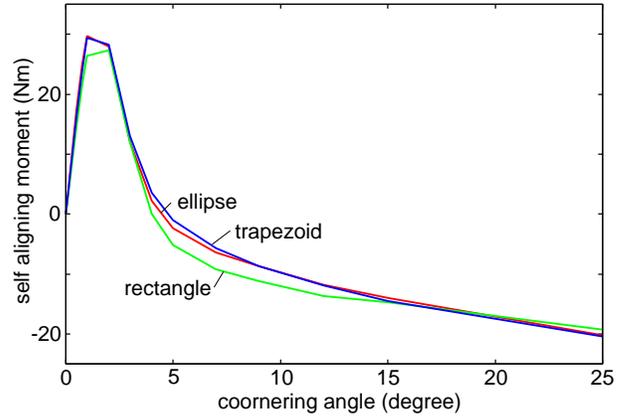}
\caption{\label{selfAligningMoment.slipangle.elips.rect.trapetz}
The self-aligning moment for the three footprints shown
in Fig. \ref{footprint} as a function of the coornering angle. 
For the rubber background temperature $T_0=80 \ ^\circ {\rm C}$
and the car velocity $27 \ {\rm m/s}$. 
For $F_{\rm N} = 2000 \ {\rm N}$ and $p = 0.1 \ {\rm MPa}$.  
}
\end{figure}

In Fig. \ref{selfAligningMoment.slip.Trapetz} we show the 
the self-aligning moment for the trapezoid footprint profile
in Fig. \ref{footprint}, as a function of the longitudinal slip (for zero cornering angle).
Note that for the rectangular and elliptic footprint the self 
aligning moment vanish (not shown).
This is expected because of the mirror symmetry of the footprint in the
$x$-axis through the center of the footprint. However, the trapezoid footprint
does not exhibit this symmetry (see Fig. \ref{footprint}), 
and the self aligning moment is non-vanishing in this case.  

In Fig. \ref{selfAligningMoment.slipangle.elips.rect.trapetz} we show the
self-aligning moment for the three footprints 
shown in Fig. \ref{footprint}, as a function of the cornering angle (for zero longitudinal slip).
In this case the self aligning moment is non-vanishing in all cases. 
It is remarkable, however,
how insensitive the results are to the shape of the footprint.

For small $\theta$ the self aligning moment is positive.
This is due to the gradual (nearly linear) build up of the transverse stress from
the inlet of the footprint to the exit. Thus the center of mass of the frictional stress distribution
is located closer to the exit of the footprint giving a positive self aligning moment.
For large $\theta$ the self aligning moment is negative. In the present model this is due to the
flash temperature effect: when a tread block enter the footprint the rubber is ``cold'',
and initially the rubber friction is high. After a short slip distance the full
flash temperature is built up and the rubber friction is smaller. This will result 
in a frictional stress distribution 
which is larger close to the inlet. 
This in turn result in a negative
self aligning moment. 

In the study above we have assumed a constant pressure in the footprints. In reality,
the pressure will be slightly larger at the inlet than at the exit of contact with the road
(this is the case also during pure rolling and is related to the rolling resistance).
This asymmetry will give an additional (negative) contribution to self aligning moment for large slip.

\begin{figure}
\includegraphics[width=0.45\textwidth,angle=0]{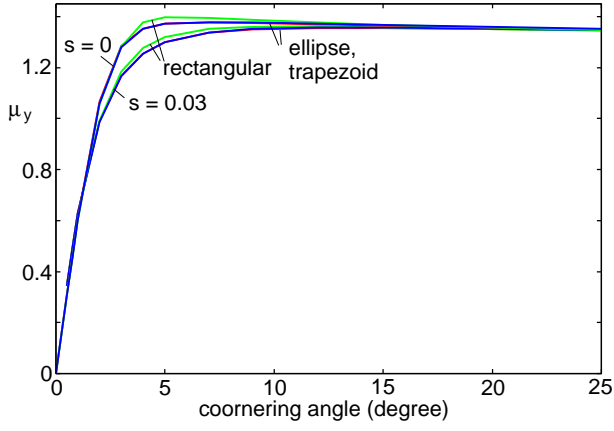}
\caption{\label{mu.slipangle.slip.0.and.0.03.elips.rect.trapetz}
The $\mu$-slip angle curves for the three footprints 
shown in Fig. \ref{footprint} and for the longitudinal slip $s=0$ and $s=0.03$. 
For the rubber background temperature $T_0=80 \ ^\circ {\rm C}$
and the car velocity $27 \ {\rm m/s}$. 
For $F_{\rm N} = 2000 \ {\rm N}$ and $p = 0.1 \ {\rm MPa}$.  
}
\end{figure}

\begin{figure}
\includegraphics[width=0.45\textwidth,angle=0]{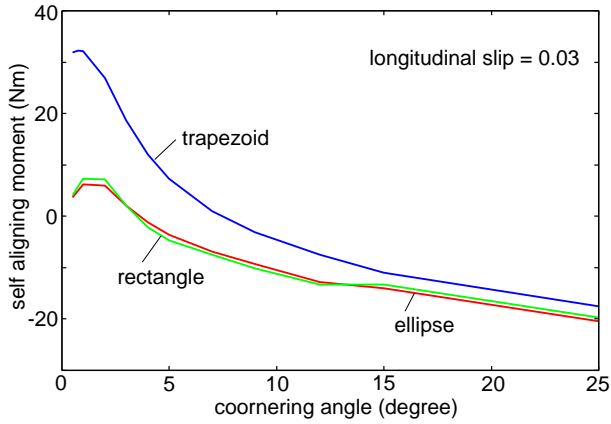}
\caption{\label{selfAligningMoment.slip.0.03.slipangle.elips.rect.trapetz}
The self-aligning moment for the three footprints
shown in Fig. \ref{footprint} and for the longitudinal slip $s=0.03$. 
For the rubber background temperature $T_0=80 \ ^\circ {\rm C}$
and the car velocity $27 \ {\rm m/s}$. 
For $F_{\rm N} = 2000 \ {\rm N}$ and $p = 0.1 \ {\rm MPa}$.  
}
\end{figure}

\begin{figure}
\includegraphics[width=0.45\textwidth,angle=0]{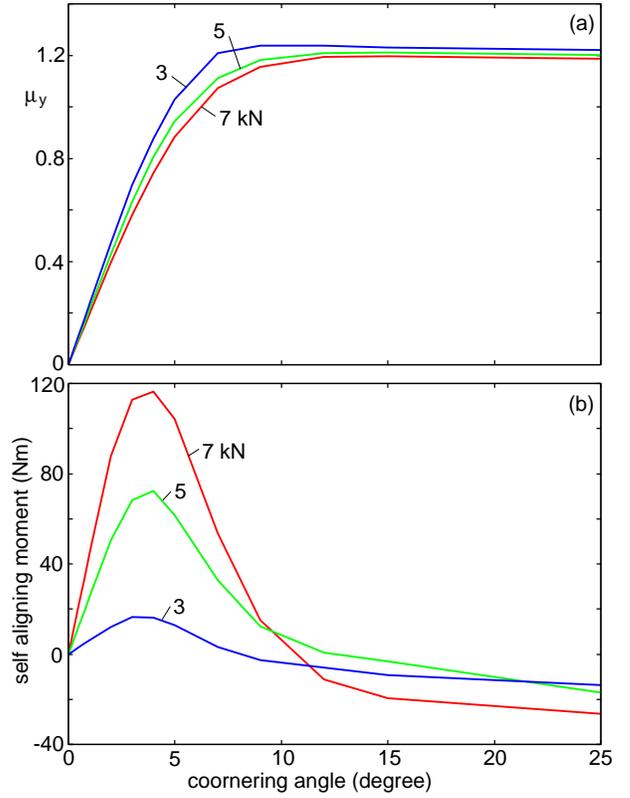}
\caption{\label{mu.slipangle.3000N.5000N.7000N.p0.3MPa.ellipse}
The $\mu$-slip angle curves (a), and the self aligning moment (b) 
for the elliptic footprint for the tire load $F_{\rm N} = 3000$, $5000$ and $7000 \ {\rm N}$,
and the footprint pressure $p=0.3  \ {\rm MPa}$. 
For the rubber background temperature $T_0=80 \ ^\circ {\rm C}$
and the car velocity $27 \ {\rm m/s}$. 
}
\end{figure}

In Fig. \ref{mu.slipangle.slip.0.and.0.03.elips.rect.trapetz} we show
the $\mu$-slip angle curves for the three footprints shown
in Fig. \ref{footprint}, and for the longitudinal slip $s=0$ and $s=0.03$. 
Note, in accordance with experimental observations, for the combined slip
$\mu_y(\theta)$ is smaller than for the case when the longitudinal slip vanish.
Again there is very little influence on the shape of the footprint. However,
the self aligning moment will now depend strongly on the shape of the 
footprint. This is shown in Fig. \ref{selfAligningMoment.slip.0.03.slipangle.elips.rect.trapetz}
for the case $s=0.03$.
Note that for the trapezoid footprint the self aligning moment is much larger than for
the elliptic and rectangular footprints. This is due to the contribution from
the longitudinal stress component which gives rise to a net longitudinal force centered to
the right of the mid-line of the tire.

Fig. \ref{mu.slipangle.3000N.5000N.7000N.p0.3MPa.ellipse} shows the 
$\mu$-slip angle curves (a), and the self aligning moment (b) 
for the elliptic footprint for the tire load $F_{\rm N} = 3000$, $5000$ and $7000 \ {\rm N}$,
and the footprint pressure $p=0.3  \ {\rm MPa}$. 
Note that as the load increases the foot print becomes longer which result in a decrease in the maximum friction
coefficient, which agree with experimental observations. 
This load-dependence is not due to an intrinsic pressure
dependence of the rubber friction coefficient (which was kept constant in our calculation), 
but a kinetic effect related to the build up of the flash
temperature in rubber road asperity contact regions during slip. To understand this in more detail, consider
again Fig. \ref{muslip}. 

The red and blue lines in Fig. \ref{muslip}
show the kinetic friction coefficient (stationary sliding) as a function of the logarithm
of the sliding velocity. The upper line denoted ``cold'' is without
the flash temperature while the lower line denoted ``hot'' is with the flash temperature.
The black curves show the effective friction experienced by a tread block  as it goes through the footprint.
Results are shown for several
slip values $0.005$, $0.0075$, $0.01$, $0.03$, $0.05$, $0.07$ and $0.09$. Note that the friction
experienced by the tread block first follows the ``cold'' rubber branch, and then, when the block has slipped a 
distance of order the diameter $D$ of the macroasperity contact region,
it follows the ``hot'' rubber branch. Based on this figure it is easy to understand why the maximum
friction coefficient increases when the length of the footprint decreases: If $v_{\rm slip}$ is the
(average) slip velocity of the tread block, then in order to fully build up the flash temperature
the following condition must be satisfied:
$v_{\rm slip} t_{\rm slip} \approx D$, where $D$ is the diameter of the macroasperity
contact region. Since the time the rubber block stay in the
footprint $t_{\rm slip} = L/v_{\rm R}$ (where $L$ is the length of the footprint and $v_{\rm R}$
the rolling velocity) we get $v_{\rm slip} \approx v_{\rm R} (D/L)$. Thus, when the length
$L$ of the footprint decreases, the (average) slip velocity of the tread block in the footprint
can increase without the slip distance exceeding the diameter $D$ of the macro asperity contact
region. Thus at the slip corresponding to the maximum of the $\mu$-slip curve, 
for a short footprint, as compared to a longer footprint, 
the tread block will follow the ``cold'' rubber branch of the (steady state) $\mu$-slip
curve to higher slip velocities before the flash temperature is fully developed, 
resulting in a higher (maximal) tire-road friction for a short 
footprint as compared to a longer footprint.  

\begin{figure}
\includegraphics[width=0.47\textwidth,angle=0]{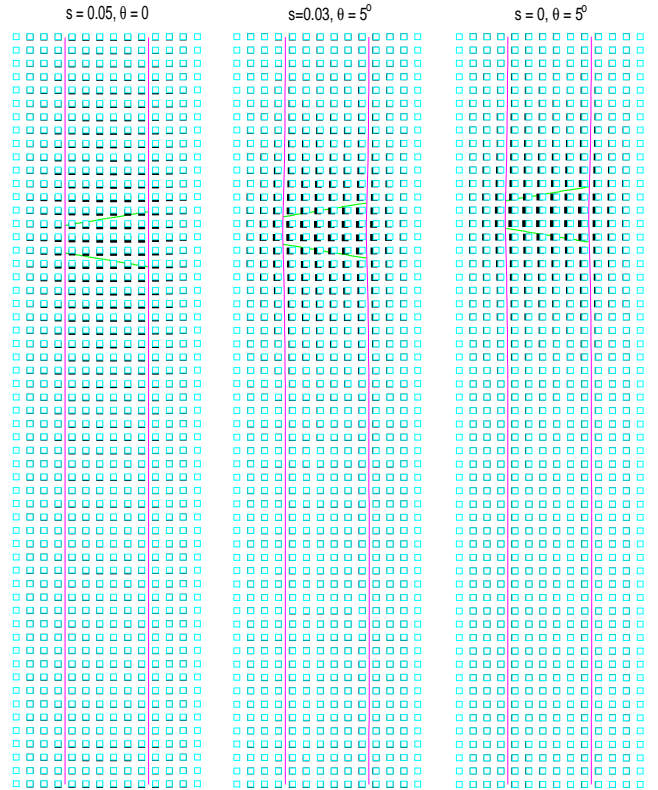}
\caption{\label{TireDeformationPicture.139}
Snapshot pictures of the tire body deformations for
slip $s$ and the coornering angle $\theta$ given by $(s,\theta) = (0.05,0)$ (left),
$(0.03, 5^\circ )$ (middle) and $(0, 5^\circ )$ (right). 
The open squares denote the position of rubber elements of the undeformed tire body,
and the filled squares underneath denote the position of the same 
tire body elements of the deformed tire.
For the trapezoid contact area
with contact pressure $0.1 \ {\rm MPa}$ and normal load $F_{\rm N} = 2000 \ {\rm N}$.
For the rubber background temperature $T_0=80 \ ^\circ {\rm C}$
and the car velocity $27 \ {\rm m/s}$. 
}
\end{figure}

\begin{figure}
\includegraphics[width=0.47\textwidth,angle=0]{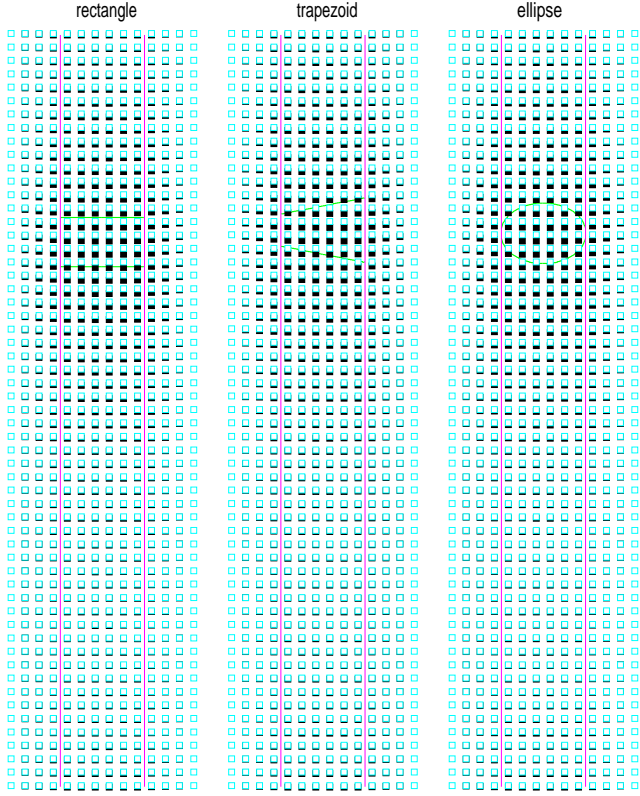}
\caption{\label{AllFootprint.7000N.TireDeformationPicture.139}
Snapshot pictures of the tire body deformations for
the rectangular, trapezoid and elliptic contact area
with contact pressure $0.3 \ {\rm MPa}$ and normal load $F_{\rm N} = 7000\ {\rm N}$.
For the slip $s=0.05$ and the coornering angle $\theta =0$. 
The open squares denote the position of rubber elements of the undeformed tire body,
and the filled squares underneath denote the position of the same 
tire body elements of the deformed tire.
For the rubber background temperature $T_0=80 \ ^\circ {\rm C}$
and the car velocity $27 \ {\rm m/s}$. 
}
\end{figure}

In Fig. \ref{TireDeformationPicture.139}
we show snapshot pictures of the tire body deformations for three cases, namely with the
slip $s$ and the coornering angle $\theta$ given by $(s,\theta) = (0.05,0)$ (left),
$(0.03, 5^\circ )$ (middle) and $(0, 5^\circ )$ (right). 
The open squares denote the position of rubber elements of the undeformed tire body,
and the filled squares underneath denote the position of the same 
tire body elements of the deformed tire.

In Fig. \ref{AllFootprint.7000N.TireDeformationPicture.139} we show
snapshot pictures of the tire body deformations for 
the rectangular, trapezoid and elliptic contact area
with contact pressure $0.3 \ {\rm MPa}$ and normal load $F_{\rm N} = 7000\ {\rm N}$.
In all cases the slip $s=0.05$ and the coornering angle $\theta =0$.
Note how insensitive the deformation field is to the shape of the footprint.
This is caused by the high stiffness of the tire body in the tread area. 

\begin{figure}
\includegraphics[width=0.45\textwidth,angle=0]{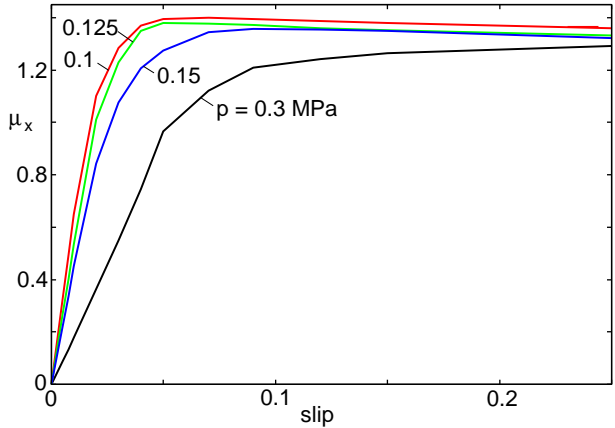}
\caption{\label{rectangular.mu.slip.pressure.1..1.25..1.5..3bar}
The $\mu$-slip curves for the rectangular footprint shown
in Fig. \ref{footprint}. For the contact pressures $p=0.1$, $0.125$, $0.15$ and $0.3 \ {\rm MPa}$
corresponding to the footprint length $L= 10.2$, $8.1$, $6.8$ and $3.4 \ {\rm cm}$. 
For the rubber background temperature $T_0=80 \ ^\circ {\rm C}$, the tire load $F_{\rm N} = 2000$,
and the car velocity $27 \ {\rm m/s}$.}
\end{figure}

\begin{figure}
\includegraphics[width=0.47\textwidth,angle=0]{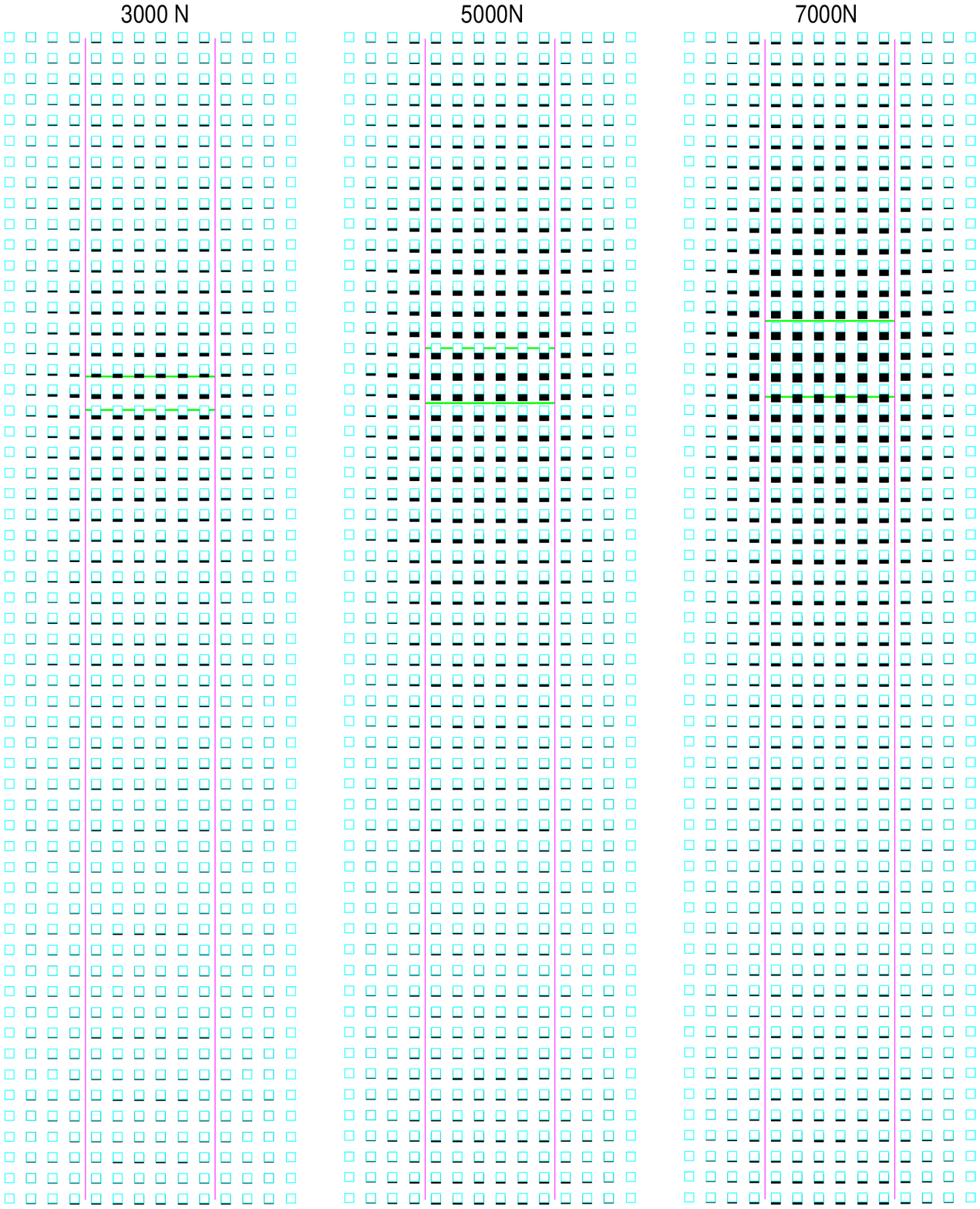}
\caption{\label{rectang.load7000N.0.3MPa.TireDeformationPicture.139}
Snapshot pictures of the tire body deformations for 
the normal load $F_{\rm N} = 3000$, $5000$  and
$7000 \ {\rm N}$.
In all cases the
slip $s=0.05$ and the coornering angle $\theta =0$. 
The open squares denote the position of rubber elements of the undeformed tire body,
and the filled squares underneath denote the position of the same 
tire body elements of the deformed tire.
The maximum tire body displacements are
$0.92$, $1.39$ and $1.84 \ {\rm cm}$ for the tire loads $F_{\rm N} = 3000$, $5000$  and
$7000 \ {\rm N}$, respectively. 
For the rectangular contact area
with contact pressure $0.3 \ {\rm MPa}$.
For the rubber background temperature $T_0=80 \ ^\circ {\rm C}$
and the car velocity $27 \ {\rm m/s}$. 
}
\end{figure}

\begin{figure}
\includegraphics[width=0.47\textwidth,angle=0]{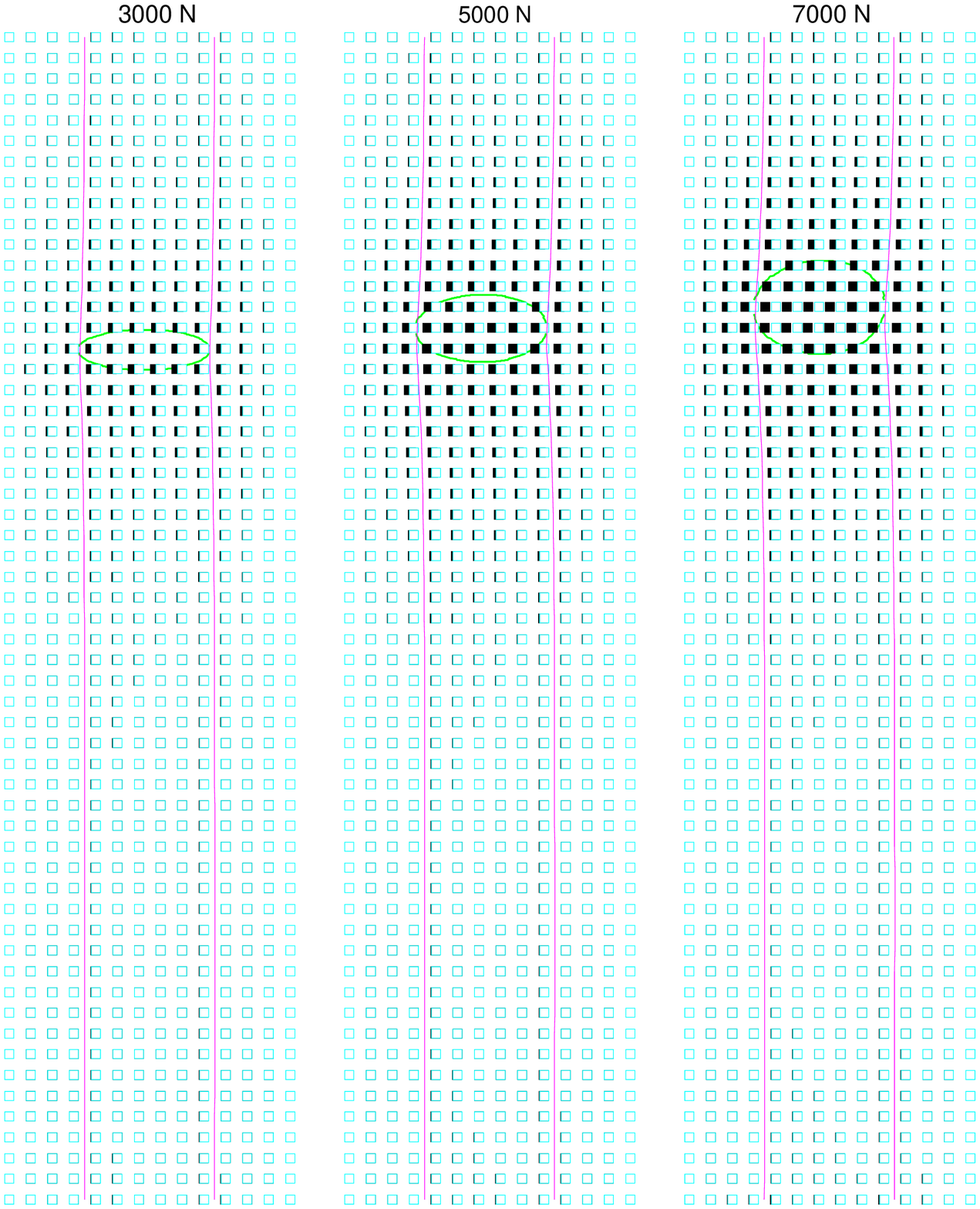}
\caption{\label{All.rect.3000.5000.7000.p0.3MPa.cornering}
Snapshot pictures of the tire body deformations for  
the normal load $F_{\rm N} = 3000$, $5000$  and
$7000 \ {\rm N}$.
In all cases the
slip $s=0$ and the coornering angle $\theta =5^\circ$. 
The open squares denote the position of rubber elements of the undeformed tire body,
and the filled squares underneath denote the position of the same 
tire body elements of the deformed tire. 
The maximum tire body displacements are
$1.20$, $1.81$ and $2.31 \ {\rm cm}$ for the tire loads $F_{\rm N} = 3000$, $5000$  and
$7000 \ {\rm N}$, respectively. 
For the elliptic contact area
with contact pressure $0.3 \ {\rm MPa}$.
For the rubber background temperature $T_0=80 \ ^\circ {\rm C}$
and the car velocity $27 \ {\rm m/s}$. 
}
\end{figure}

\begin{figure}
\includegraphics[width=0.45\textwidth,angle=0]{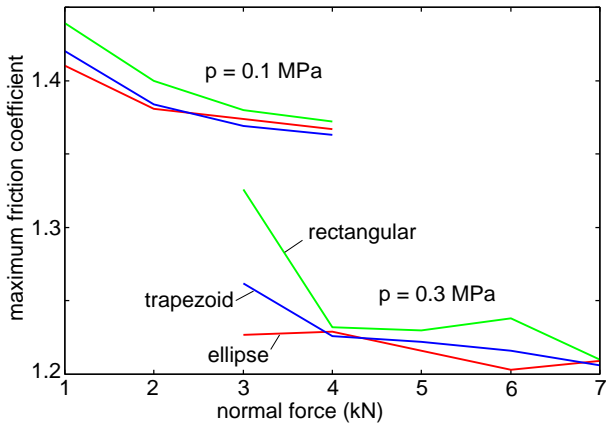}
\caption{\label{elips.rect.trapezoid.FN.mumax.0.1MPa.0.3MPa}
The maximum friction coefficient, $\mu_{\rm max}$, of the $\mu_x$-slip curve for
elliptic, rectangular and trapezoid footprints.
For the contact pressures $p=0.1 \ {\rm MPa}$ (upper three curves)
and $0.3 \ {\rm MPa}$ (lower three curves).
For the rubber background temperature $T_0=80 \ ^\circ {\rm C}$
and the car velocity $27 \ {\rm m/s}$. 
}
\end{figure}

\begin{figure}
\includegraphics[width=0.45\textwidth,angle=0]{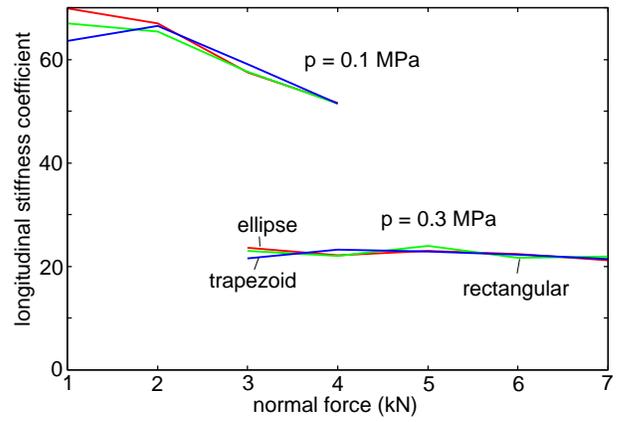}
\caption{\label{longitudinalStiffness.ellips.rectangle.trapezoid}
The longitudinal tire stiffness $C_{\rm L}$ associated with the $\mu_x$-slip curve for 
elliptic, rectangular and trapezoid footprints. 
For the contact pressures $p=0.1 \ {\rm MPa}$ (upper three curves)
and $0.3 \ {\rm MPa}$ (lower three curves).
For the rubber background temperature $T_0=80 \ ^\circ {\rm C}$
and the car velocity $27 \ {\rm m/s}$. 
}
\end{figure}

\begin{figure}
\includegraphics[width=0.45\textwidth,angle=0]{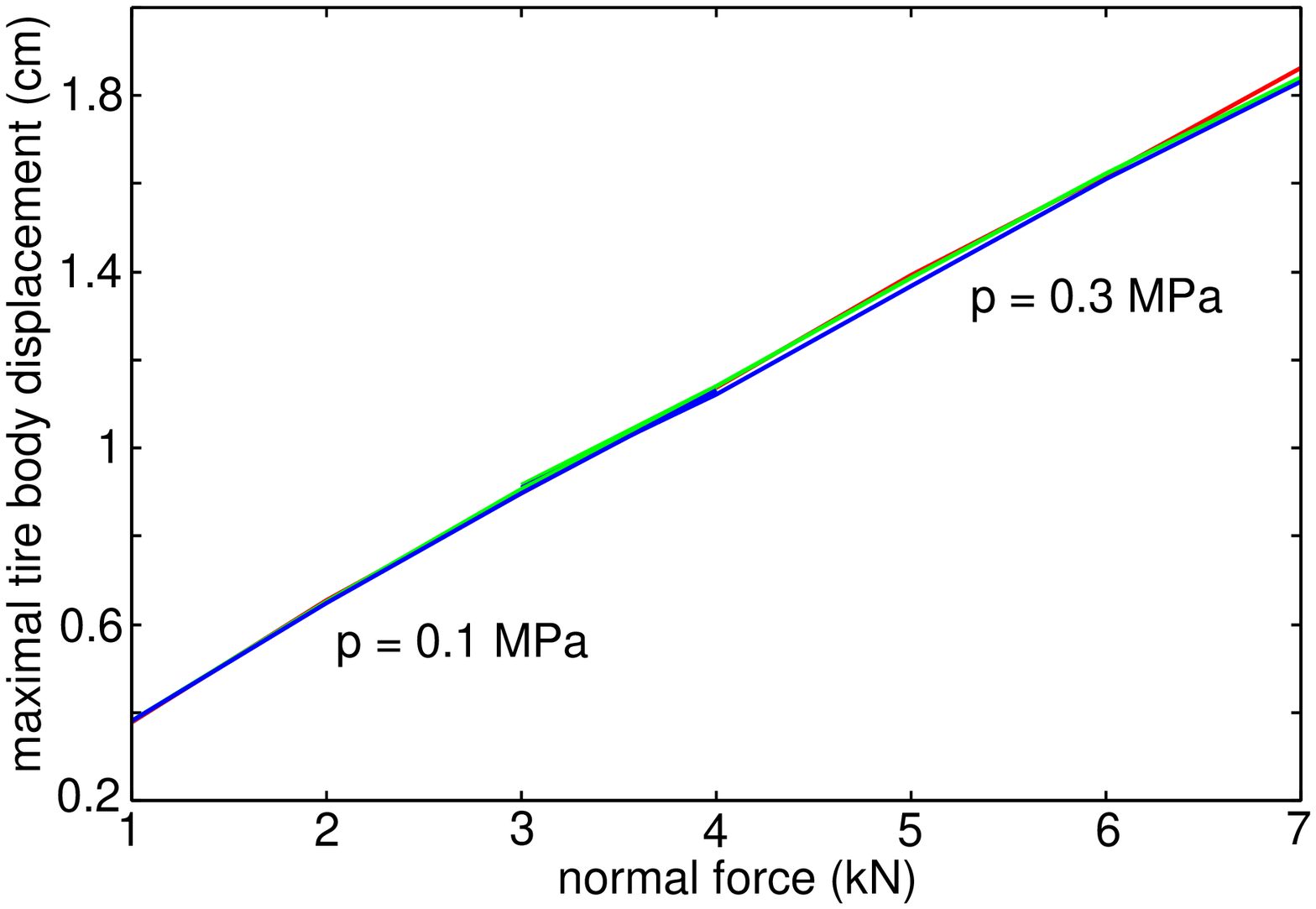}
\caption{\label{displacement.longitudinal.all}
The maximum longitudinal tire body displacement for 
elliptic, rectangular and trapezoid footprints.
For the contact pressures $p=0.1 \ {\rm MPa}$ (lower three curves)
and $0.3 \ {\rm MPa}$ (upper three curves).
For the rubber background temperature $T_0=80 \ ^\circ {\rm C}$
and the car velocity $27 \ {\rm m/s}$. 
}
\end{figure}

\vskip 0.2cm
{\bf 4.2 Dependence of the $\mu$-slip curve on the size of the tire-road footprint}

In Fig. \ref{rectangular.mu.slip.pressure.1..1.25..1.5..3bar} we show $\mu$-slip curves for the rectangular footprint
for the contact pressures $p=0.1$, $0.125$, $0.15$ and $0.3 \ {\rm MPa}$. The tire load is fixed at $F_{\rm N} = 2000$ so
the different contact pressures
corresponding to the footprint length $L= 10.2$, $8.1$, $6.8$ and $3.4 \ {\rm cm}$, respectively. 
Note that increasing the tire footprint pressure decreases the length of the footprint, which decreases
the tire longitudinal stiffness $C_x$ (determined by the initial slope of the $\mu_x(s)$-curves), 
and also the maximum of the $\mu$-slip curves. 
In practice the pressure in the tire-road footprint
can be changed by changing the tire inflation pressure, but in this case one expect
also a change in the tire body stiffness.
In the tire model we use this effect
is not included at present. Thus, at least for passenger car tires, when the inflation pressure increases the longitudinal
tire stiffness $C_x$ usually first increases and then, at high enough inflation pressure, decreases. This is consistent
with Fig. \ref{rectangular.mu.slip.pressure.1..1.25..1.5..3bar} which shows initially a very small change in the tire
stiffness as the footprint pressure $p$ increases, so that for small (but not too small) $p$, 
the stiffening of the tire body may dominate over the contribution from the change in the footprint, so that initially
$C_x$ increases with increasing $p$.

In Fig. \ref{rectang.load7000N.0.3MPa.TireDeformationPicture.139}
we show snap shot pictures
of the tire body deformations for three cases, namely 
for the external load
$F_{\rm N} = 3000$, $5000$  and
$7000 \ {\rm N}$.
In all cases the
slip $s=0.05$ and the coornering angle $\theta =0$. 
The results are for the rectangular footprint
with contact pressure $0.3 \ {\rm MPa}$.

In Fig. \ref{All.rect.3000.5000.7000.p0.3MPa.cornering}
we show similar results as in Fig. \ref{rectang.load7000N.0.3MPa.TireDeformationPicture.139}
but now for the elliptic contact area with contact pressure $0.3 \ {\rm MPa}$, and with
the slip $s=0$ and the coornering angle $\theta =5^\circ$. 
The maximum tire body displacements are
$1.20$, $1.81$ and $2.31 \ {\rm cm}$ for the tire loads $F_{\rm N} = 3000$, $5000$  and
$7000 \ {\rm N}$, respectively. 

In Fig. \ref{elips.rect.trapezoid.FN.mumax.0.1MPa.0.3MPa}
we show the maximum friction coefficient, $\mu_{\rm max}$, of the $\mu_x$-slip curve for the 
elliptic, rectangular and trapezoid footprints, as a function of the tire load. 
We show results for the contact pressures $p=0.1 \ {\rm MPa}$ (upper three curves)
and $0.3 \ {\rm MPa}$ (lower three curves), but the tire body properties are assumed to be the same.
Note that the effective friction depends on the (average) tire-road footprint pressure $p$.
When $p$ increases, assuming unchanged size of the footprint, the friction decreases.
This is one reason for why racer tires exhibit much larger friction than passenger car tires 
(the contact pressure in F1 tires is of order $0.1 \ {\rm MPa}$ which is about 3 times lower than in passenger car tires).
The increase in the friction as the tire-road contact pressure $p$ decreases is mainly due to a decrease in the
pressure in the macro asperity contact regions as $p$ decreases: When the local pressure decreases
the produced heating of the rubber decreases leading to a larger friction.

In Fig. \ref{longitudinalStiffness.ellips.rectangle.trapezoid} we show for the same systems
the longitudinal tire stiffness $C_x$ associated with the $\mu_x$-slip curves.
Finally, Fig. \ref{displacement.longitudinal.all} gives the maximum longitudinal tire body displacement 
as a function of the tire load, for the same systems as in Fig. \ref{elips.rect.trapezoid.FN.mumax.0.1MPa.0.3MPa}.

\vskip 0.2cm
{\bf 4.3 Dependence of the $\mu$-slip curve on the car velocity}

\begin{figure}
\includegraphics[width=0.45\textwidth,angle=0]{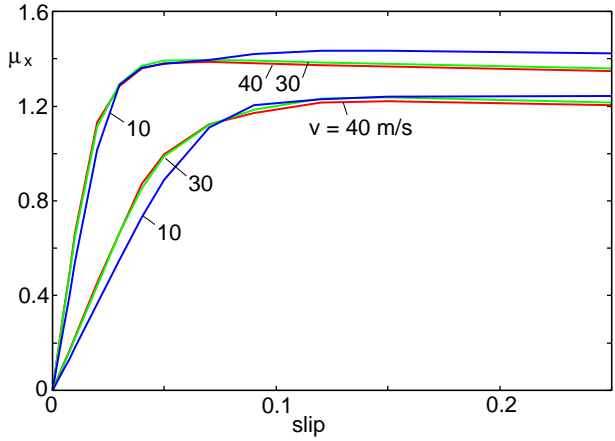}
\caption{\label{slip.mu.rectangular.foot.40ms.30ms.10ms}
The $\mu$-slip curves for a rectangular footprint ($20 \ {\rm cm}\times 10.2 \ {\rm cm}$)
and for the car velocities $v_{\rm c} = 10$, $30$ and $40 \ {\rm m/s}$.
For $F_{\rm N} = 2000 \ {\rm N}$ and $p = 0.1 \ {\rm MPa}$ (top three curves)
and $F_{\rm N} = 6000 \ {\rm N}$ and $p = 0.3 \ {\rm MPa}$ (lower three curves).  
For the rubber background temperature $T_0=80 \ ^\circ {\rm C}$. 
}
\end{figure}

Fig. \ref{slip.mu.rectangular.foot.40ms.30ms.10ms} shows the 
the $\mu$-slip curves for the rectangular footprint ($20 \ {\rm cm}\times 10.2 \ {\rm cm}$)
shown in Fig. \ref{footprint}, and for the car velocities $v_{\rm c} = 10$, $30$ and $40 \ {\rm m/s}$.
We show results both for the tire load $F_{\rm N} = 2000 \ {\rm N}$ and $p = 0.1 \ {\rm MPa}$ (top three curves)
and for $F_{\rm N} = 6000 \ {\rm N}$ and $p = 0.3 \ {\rm MPa}$ (lower three curves).  
Note that as the car velocity decreases the longitudinal tire stiffness $C_x$ decreases and the maximum
of the $\mu$-slip curve increases. This is in accordance with experimental observations for passenger car
tires (see, e.g., Fig. 8.65 in Ref. \cite{tire}). When the contact pressure and the load both increases, 
in such a way that the contact area stays constant, both the tire stiffness and the maximum of the
$\mu$-slip curves decreases.

\begin{figure}
\includegraphics[width=0.45\textwidth,angle=0.0]{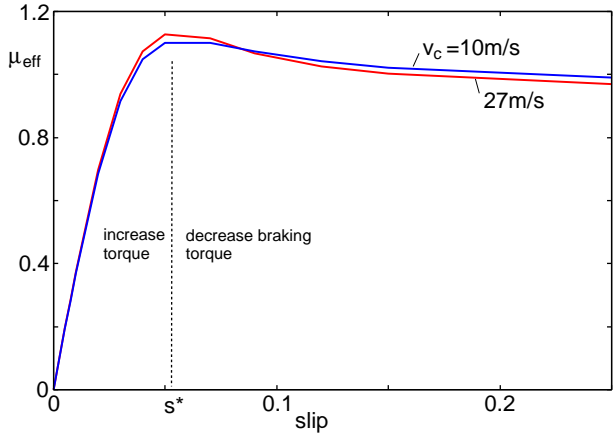}
\caption{\label{mu.slip.10ms.27ms}
The $\mu$-slip curves for the car velocity $v_{\rm c}=10$ and $27 \ {\rm m/s}$. 
The maximum $\mu^*$ of the $\mu$-slip curve, and the slip $s=s^*$ where the maximum occur, 
depends on the car velocity $v_{\rm c}$.
In the present case $s^*\approx 0.057$  for both velocities.
The ABS control algorithm should increase the braking torque when $s < s^*$ and
reduce the braking torque when $s > s^*$. 
}
\end{figure}

\vskip 0.3cm
{\bf 5. ABS braking simulations}

The theory developed above may be extremely useful to design or optimize control
algorithms for ABS braking. Here we will present results using the two simplest possible
control algorithms. In both cases the braking torque is changed (increase or decrease) 
in steps of $\Delta M$ at time $t_n= n \Delta t$ $(n=1,2,...)$. 
The first algorithm (a) assumes that the slip $s^*$ 
where the friction is maximal is known (and constant in time). 
In this case the braking torque is increased if the slip
$s(t_n)$ at time $t_n$ is below $s^*$ and otherwise it is decreased, see Fig. \ref{mu.slip.10ms.27ms}. 
One problem here is that the slip $s^*$ depends on the car velocity which change during the braking
process. However, in the present case $s^* \approx 0.057$ nearly independent of the 
car velocity for $10 \ {\rm m/s} < v_{\rm c} < 27 \ {\rm m/s}$.

In the second control algorithm (b) we assume that $s^*$ is unknown. 
Nevertheless, by registering if the (longitudinal) friction $F_x(t)$ is increasing or decreasing with time
we can find out if we are to the left or right of the maximum at $s=s^*$.
That is, if 
$$F_x(t_{n}) > F_x(t_{n-1}) \ \ \ \ {\rm and} \ \ \ \ s(t_{n}) < s(t_{n-1})$$
or if
$$F_x(t_{n}) < F_x(t_{n-1}) \ \ \ \ {\rm and} \ \ \ \ s(t_{n}) > s(t_{n-1})$$
then we must have $s(t_n) > s^*$ and the braking torque at time $t_n$ is reduced, 
otherwise it is increased.
Here $F_x(t_n)$ is the longitudinal friction force and $s(t_n)$ the slip at time
$t_n= n \Delta t$ $(n=1,2,...)$. 

We now present numerical results to illustrate the two
ABS braking algorithm.
Let $M$ be the mass-load on a wheel and $I$ the moment of inertia of the
wheel {\it without} the tire. 
We assume for simplicity that the suspension is rigid and neglect mass-load transfer. 
The equations of motion for the center of mass coordinate $x(t)$ of the wheel, and for the
angular rotation coordinate $\phi (t)$ are:
$$M \ddot x = F_{\rm rim}\eqno(2)$$
$$I\ddot \phi = M_{\rm rim} -M_{\rm B}\eqno(3)$$
where $F_{\rm rim}$ is the force acting on the rim,
$M_{\rm B}$ is the braking torque and $M_{\rm rim}$ the torque acting on the rim from the
tire (for constant rolling velocity $F_{\rm rim} = F_{\rm f}$ is the tire-road friction force
and $M_{\rm rim} = R F_{\rm f}$, where $R$ is the rolling radius, 
but during angular accelerations
these relation no longer holds because of tire inertia effects)
We have used $M=360 \ {\rm kg}$ and $I=0.4 \ {\rm kgm^2}$. 

\begin{figure}
\includegraphics[width=0.45\textwidth,angle=0.0]{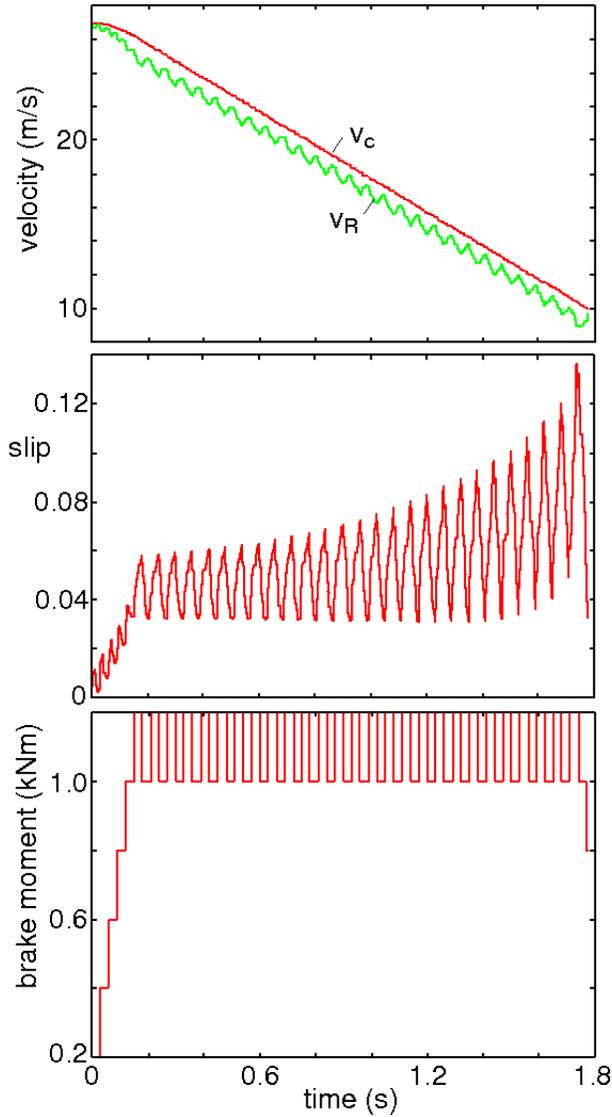}
\caption{\label{time.vc.vR.FAST}
(a) The car velocity $v_{\rm c}$ and the rolling velocity $v_{\rm R}$ 
as a function of time $t$.
(b) the slip and (c) the braking moment
as a function of time $t$. For ABS braking 
using algorithms {\bf a} (see text for details). 
}
\end{figure}

We assume first the control algorithm (a). We take $\Delta M = 200 \ {\rm Nm}$ and
$\Delta t = 0.03 \ {\rm s}$, and we assume (see Fig.
\ref{mu.slip.10ms.27ms}) $s^* \approx 0.05$.
In Fig. \ref{time.vc.vR.FAST} we show 
(a) the car velocity $v_{\rm c}$ and the rolling velocity $v_{\rm R}$, (b) the longitudinal 
slip and (c) the braking torque as a function of time.
The time it takes ($t=1.775 \ {\rm s}$) to reduce the car velocity from $v_0 = 27$ to $v_1 = 10 \ {\rm m/s}$
correspond to a friction coefficient $\mu = (v_0-v_1)/gt =0.976$ which is $\sim 13 \%$ smaller than the friction at the
maximum of the $\mu$-slip curve, which varies between $\mu_{\rm max} = 1.14$ and $1.11$ as the car velocity changes
from $27$ to $10 \ {\rm m/s}$. 
The slope of the car-velocity line in Fig. \ref{time.vc.vR.FAST}(a) for $t>0.2 \ {\rm s}$
correspond to the friction coefficient $1.02$, which is larger than the (average) friction calculated from the stopping time. 
The slightly smaller friction obtained from the stopping time reflect the (short)
initial time interval necessary to build up the braking torque.

\begin{figure}
\includegraphics[width=0.45\textwidth,angle=0.0]{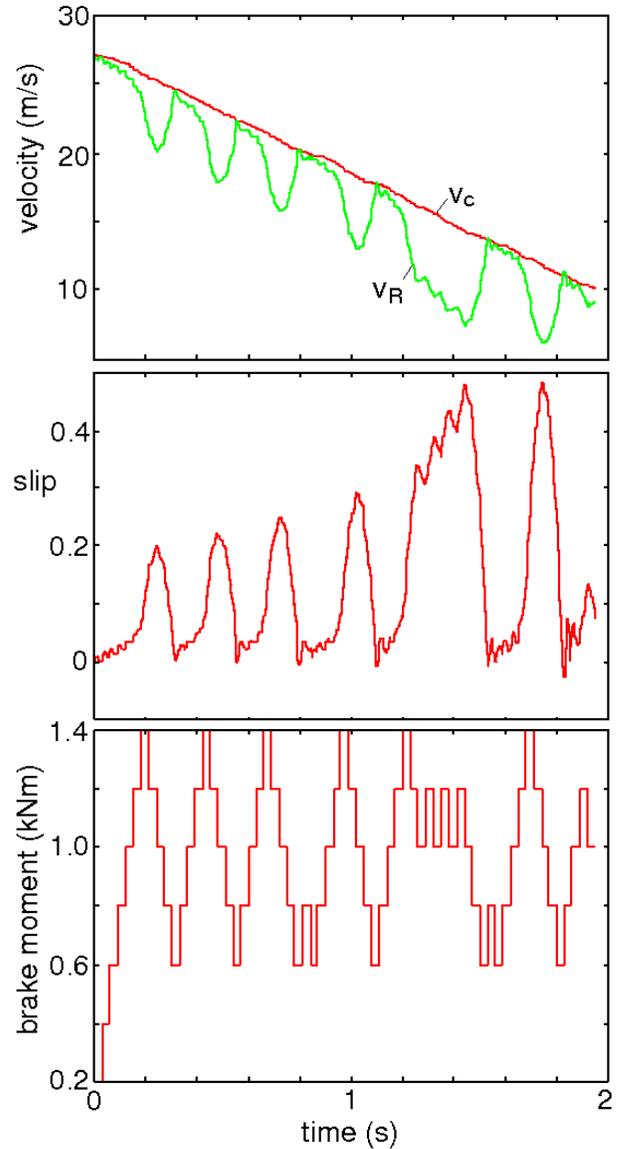}
\caption{\label{secondABS.time.vc.vR}
(a) The car velocity $v_{\rm c}$ and the rolling velocity $v_{\rm R}$ 
as a function of time $t$.
(b) the slip and (c) the braking moment
as a function of time $t$. For ABS braking 
using algorithms {\bf b} (see text for details). 
}
\end{figure}

In Fig. \ref{secondABS.time.vc.vR}
we show results for the ABS control algorithm (b).
Note that it takes $t=1.949 \ {\rm s}$ to reduce the 
car velocity from
$v_0=27 \ {\rm m/s}$ to $v_1=10 \ {\rm m/s}$. This correspond to the effective friction coefficient
$\mu = (v_0-v_1)/gt = 0.889$. The maximum in the $\mu$-slip curve (see Fig. \ref{mu.slip.10ms.27ms}) 
depends on the car velocity but is about $\mu_{\rm max} = 1.14$ for $v_{\rm c} = 27 \ {\rm m/s}$ and 
about $1.11$ for $v_{\rm c} = 10 \ {\rm m/s}$
so the ABS braking control algorithm used above could still be improved.
Note also that the wheel tend to lock about 3 or 4 times per second. This is in good agreement with 
ABS braking systems presently in use. However, since the speed of cars are usually not known during
ABS braking, braking control algorithms used in most cars today determine the braking torque only from
the wheel rotation acceleration. This is possible because, as shown in Fig. \ref{secondABS.time.vc.vR}(a),
as the wheel tend to lock, the rotational velocity very rapidly decreases, and it this point the ABS system
decreases the braking torque. 

Note that the (average) of the slip in Fig. \ref{time.vc.vR.FAST}(b) and \ref{secondABS.time.vc.vR}(b) increases with increasing
time or, equivalently, decreasing car velocity. This is mainly due to the fact that 
the time it takes for the wheel to lock, when the slip $s > s^*$, decreases as $v_{\rm c}$ decreases. Thus, during the time
period $\Delta t$ between two changes in the brake torque the maximal slip (corresponding to the minimal rolling
velocity) will increase as $v_{\rm c}$ decrease. This is easy to show mathematically. Since the car velocity changes slowly 
compared to the rolling velocity, from the definition
$s=(v_{\rm c}-v_{\rm R})/v_{\rm c}$ we get
$${d v_{\rm R} \over dt} \approx - v_{\rm c} {ds \over dt}$$
If we approximate the $\mu$-slip curve for $s > s^*$ with a strait line,
$$\mu_{\rm eff} \approx \mu_0 - \Delta \mu s,$$
we get from (3)
$$I {d^2 \phi\over dt^2} = {I \over R} {d v_{\rm R} \over dt} \approx - {I v_{\rm c}\over R} {ds \over dt} = Mg[\mu_0 - \Delta \mu s] -M_{\rm B}$$
or
$${ds \over dt} = -A+Bs$$
where $A=(MgR\mu_0- M_{\rm B})(R/Iv_{\rm c})$ and $B= \Delta \mu (MgR^2/I v_{\rm c})$.
Since $A$ and $B$ can be considered as constant during the time interval between the changes in the
braking torque, we get
$$s(t) = \left (s(0)-{A\over B}\right ) e^{Bt} +{A\over B}$$
where 
$${A\over B} = {1\over \Delta \mu} \left (\mu_0-{M_{\rm B}\over MgR}\right ).$$ 
It is easy to show that
$$s(0)-{A\over B} = [s(0)-s^*] + {M_{\rm B}^* - M_{\rm B}\over \Delta \mu M g R}$$
where $M_{\rm B}^* = Mg (\mu_0 - \Delta \mu s^*)$ is the braking torque necessary in order to stay at the maximum
in the $\mu$-slip curve. If $M_{\rm B} <M_{\rm B}^*$ and $s(0) > s^*$ we have $s(0)-A/B > 0$ and
during the time interval $\Delta t$ the slip will increase with $[s(0)-A/B] {\rm exp}(B\Delta t)$. Since $B\Delta t \sim 1/v_{\rm c}$
the maximum slip will increase exponentially (until the wheel block, corresponding to $s=1$) with the inverse of the
car velocity.  
This behavior (i.e., the increase in the slip with decreasing car velocity) can be seen in Fig. \ref{time.vc.vR.FAST}(b) and is even stronger for the second ABS
control algorithm [fig. \ref{secondABS.time.vc.vR}(b)].

\begin{figure}
\includegraphics[width=0.45\textwidth,angle=0.0]{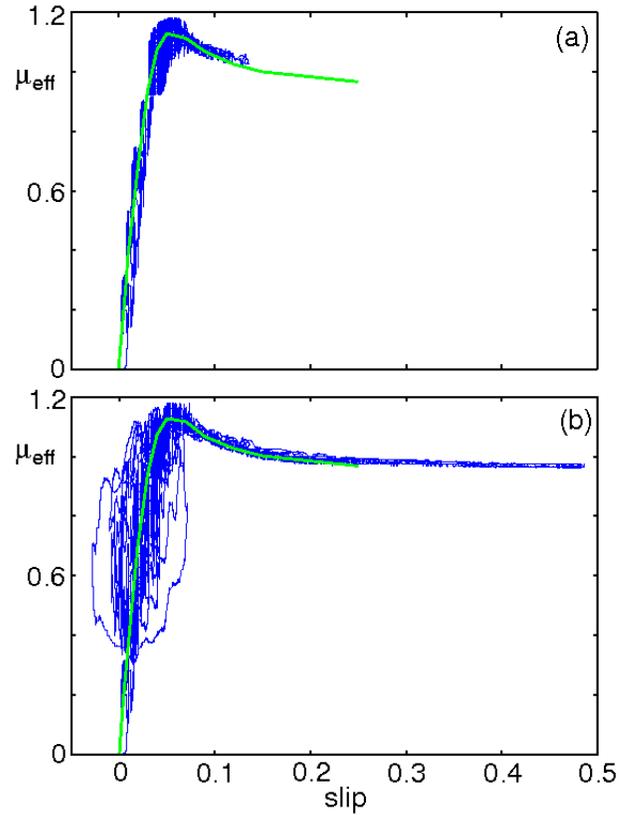}
\caption{\label{dynamical.mu.slip.both.method.ps}
Dynamical $\mu$-slip curves for
ABS braking using algorithms {\bf a} (top)  and {\bf b} (bottom).
The green curve is the steady-state $\mu$-slip curve for the car velocity $v_{\rm c} = 27 \ {\rm m/s}$.}
\end{figure}

In Fig. \ref{dynamical.mu.slip.both.method.ps}
we show
the $\mu$-slip curve during stationary slip (green curve) and the
instantaneous effective friction coefficient $\mu_{\rm eff}(t) = F_x(t)/F_{\rm N}$ during
braking (blue curve). 

\begin{figure}
\includegraphics[width=0.45\textwidth,angle=0.0]{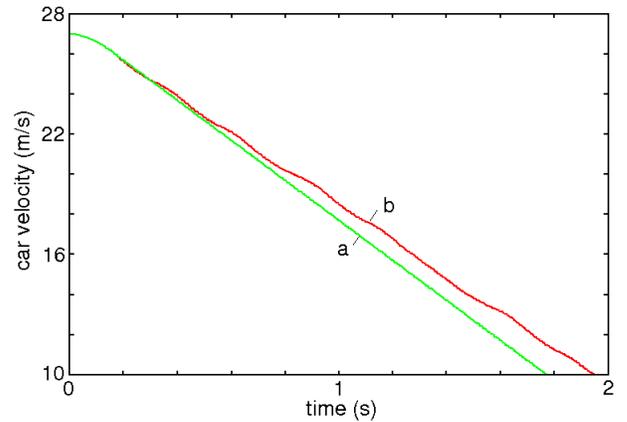}
\caption{\label{time.vcar.both}
The car velocity $v_{\rm c}$ as a function of time $t$ during ABS braking
using two algorithms {\bf a} and {\bf b}. The procedures {\bf a} and {\bf b}
results in the time periods $1.76$ and $1.95 \ {\rm s}$ for reducing the car
velocity from $27$ to $10 \ {\rm m/s}$. The effective friction values
0.976 and 0.889 are both smaller than the maximum kinetic friction  
which is 1.098 and occur at the slip velocity $0.0316 \ {\rm m/s}$. 
}
\end{figure}

The blue and red curves in Fig. \ref{time.vcar.both} show the 
car velocity using the ABS control algorithms (a) and (b).
It is clear that the ABS control algorithm (a) is more effective than algorithm (b), but  
algorithm (a) assumes
that $s^*$ is known and remains constant during the braking event.

The ABS braking control algorithms used today usually assumes that {\it only} the wheel rolling
velocity $v_{\rm R}(t)$ is known. Basically, whenever a wheel tend to lock-up, which manifest
itself in a large (negative) wheel angular acceleration, the braking torque is reduced.
These ABS braking control algorithms (e.g., the Bosch algorithm) are rather complex and secret.
The calculations presented above can be easily extended to such realistic ABS braking control algorithms
and to more complex cases such as
braking during load fluctuations (e.g., braking on uneven road surfaces) and switching 
between different road surfaces (by using different road surface power spectra's during an
ABS braking simulation).

\vskip 0.3cm
{\bf 6. Summary and conclusion}

In this paper we have proposed a simple rubber friction law, which can be used, e.g., in models of 
tire (and vehicle) dynamics. The friction law gives nearly the same result as the 
full rubber friction theory of Ref. \cite{JPCM}, but is much more convenient to use
in numerical studies of, e.g., tire dynamics, as the friction force can be calculated much faster. 

We have proposed a two-dimensional (2D) tire model which combines the 
rubber friction law with a simple mass-spring
description of the tire body. The tire model is very flexible and can be used to
calculate accurate $\mu$-slip (and the self-aligning torque) curves 
for braking and cornering or combined motion (e.g., braking during cornering). We 
have presented numerical results which illustrate the theory. Simulations of
Anti-Blocking System (ABS) braking was performed using two simple control algorithms.

\end{document}